\documentclass[11pt,twoside]{article}
\usepackage{asp2010}
\usepackage{graphicx}

\resetcounters

\begin{document}

\title{Amplitude variation and multiplet structures: Is PG\,1605+072 a slow rotator?}

\author{Jan~Langfellner,$^1$ Sonja~Schuh,$^1$ The MSST and WET teams
\affil{$^1$Georg-August-Universit\"at~G\"ottingen, Institut~f\"ur~Astrophysik, Friedrich-Hund-Platz~1, D--37077~G\"ottingen, Germany}}

\begin{abstract}
The subdwarf B star PG\,1605+072, with an unusually low log\,($g$/cm\,s$^{-2})$ $\sim 5.3$, shows a high number of non-radial pulsation modes, making it a promising candidate for asteroseismology. This could allow probing the star's interior to gain important insights in its structure and evolution. Comparison of previous work conducted over the last decade shows clear amplitude variation and hints of frequency variation. We analyse white light photometric data of the Multi-Site Spectroscopic Telescope (MSST) and Whole Earth Telescope (WET) XCov22 campaigns using prewhitening techniques and O$-$C diagrams. A total of 85 significant frequencies are identified, among them more than 20 frequency sums and harmonics. Moreover, it is shown that the main mode's amplitude varies like a sine with a period of $\sim 630\,$d, indicative of long-term beating. Strong hints for the existence of frequency multiplets support the hypothesis that PG\,1605+072 is a slow rotator with $V_\mathrm{eq} \lesssim 0.9\,\mathrm{km}\,\mathrm{s}^{-1}$, contrary to previous claims. Existing asteroseismic models mainly suggest that the star possesses a high mass of $\sim 0.7\,M_\odot$, presuming the main pulsation mode to be radial. This calls for an alternate formation channel.
\end{abstract}

\section{Introduction}	\label{introduction_section}
The subdwarf B star PG\,1605+072, also known as V338\,Ser, was found to be a $p$-mode pulsator by \citet{koen_1998}, showing more than 30 pulsation frequencies. The main mode at 2076\,mHz ($\sim 179\,$c$\,$d$^{-1}$) initially proved to be very strong, possessing an amplitude of more than 60\,mmi. This rich pulsation spectrum was confirmed by the two-week photometric multi-site campaign that was conducted by \citet{kilkenny_1999} in 1997, revealing a total of 55 pulsation frequencies, with periods ranging from 200\,s to 600\,s. However, the authors found that the main mode's amplitude decreased to less than 30\,mmi on the timescale of roughly one year. The analysis of spectroscopic and photometric data by \citet{otoole_2000}, obtained in 1999, showed that the main mode not only weakened, but completely disappeared. In spectroscopic campaigns conducted in 2000 by \citet{otoole_2002} and \citet{woolf_2002}, the mode returned but was still weaker than in 1997. Finally, spectroscopic and photometric data from 2001 by \citet{falter_2003} produced a spectrum similar to the findings in \citet{kilkenny_1999}. This behaviour hints at amplitude variation on the timescale of years.

Spectroscopically, PG\,1605+072 shows $T_\mathrm{eff} = 32\,100 \pm 1\,000$\,K, which is typical of a $p$-mode pulsating sdB star, but also a low log\,($g$/cm\,s$^{-2}) = 5.25 \pm 0.10$ \citep[confirmed by later observations]{koen_1998}, putting the star in an unusual position in the Hertzsprung-Russell diagram, about 0.6 dex away from the zero-age extreme horizontal branch. This can either be explained by an evolved status \citep{dorman_1993} or an unusually high mass \citep[see, e.\,g.,][]{vangrootel_2010}.

Another problem which has not been solved to this day is the rotation velocity. \citet{heber_1999} considered the line broadening observed in PG\,1605+072 to be caused by rotation, delivering a rotation velocity of $39\,\mathrm{km}\,\mathrm{s}^{-1}$. This value is unusually high for an sdB star, especially considering that PG\,1605+072 seems to be a single star: It will evolve to a single white dwarf and keep its high rotation velocity (if there is no mechanism to get rid of the angular momentum), but single white dwarfs are measured to be slow rotators \citep[and references therein]{heber_1997}. \citet{geier_2010} also measure rather low rotational velocities for sdB stars, namely $\sim 8\,\mathrm{km}\,\mathrm{s}^{-1}$ for the majority of their targets. On the other hand, a high rotation velocity would be able to explain the rich pulsation spectrum, as it lifts the $(2l+1)$-fold degeneracy of pulsation modes with the same $k$ and $l$ but different $m$ values. Thus, \citet{kawaler_1999} interpreted the spectrum as mainly being the result of trapped non-radial modes. The authors estimated a rotation velocity of $\sim 130\,\mathrm{km}\,\mathrm{s}^{-1}$.

However, the line broadening can also be interpreted as pulsational broadening. \citet{kuassivi_2005} came to the conclusion that the rapid Doppler shift of the lines due to pulsation got smeared in the data of \citeauthor{heber_1999}\ because of the long integration time of 600\,s. Using FUSE (Far Ultraviolet Spectroscopic Explorer) data, they found a Doppler shift of $\sim 17\,\mathrm{km}\,\mathrm{s}^{-1}$. The order of magnitude is consistent with the results of \citet{otoole_2002} and \citet{woolf_2002} as well as the results from the spectroscopic part of the Multi-Site Spectroscopic Telescope (MSST) campaign \citep{otoole_2005}. The complexity of the pulsation spectrum could be explained by assuming the low-amplitude modes to be second- and third-order harmonics as well as non-linear combinations of the very strong main modes. Following this idea, the pulsation spectrum from \citet{kilkenny_1999} can be built up using just 22 basic frequencies \citep{vangrootel_2010}.

We investigate these problems using photometric data from the MSST and Whole Earth Telescope (WET) XCov22 campaigns (see Sect.~\ref{data_section}). In Sect.~\ref{prewhitening_section}, the results from the prewhitening process are presented. Evidence for amplitude variation is followed in Sect.~\ref{amp-variation_section} and mode splitting is analysed in Sect.~\ref{multiplet_section}. The implications for our understanding of PG\,1605+072 are then discussed in Sect.~\ref{discussion_section}, before the conclusions are drawn in Sect.~\ref{conclusions_section}.

\section{Data}	\label{data_section}
PG\,1605+072's rich frequency spectrum, high amplitudes and long pulsation periods were considered excellent prerequisites for detecting spectroscopic variations and applying asteroseismic methods to understand its evolutionary status. Thus, in 2002, the Multi-Site Spectroscopic Telescope (MSST) campaign was conducted by \citet{heber_2003} in order to obtain simultaneous high quality spectroscopic and photometric data, allowing the resolution of more pulsation frequencies without strong aliasing effects. Additionally, the MSST project was supported by the selection of PG\,1605+072 as an alternate target in the Whole Earth Telescope (WET) XCov22 photometric campaign \citep{schuh_2003}.

Parts of the campaign data have already been analysed \citep{otoole_2004}. We analysed the combined photometric data, spanning a time base of approximately 30 days. The time stamps have been converted into BJD$_\mathrm{TDB}$. Since many individual light curves from different sites overlapped in time, cross-cor\-re\-la\-tion could be used to uncover and if necessary eliminate remaining timing errors. Overlapping light curve fragments with timestamps known to be already in the correct format were used to create a model light curve in order to cover the whole time base. The remaining data could be compared to this model using cross-cor\-re\-la\-tion again. This way, timing problems could be identified and treated individually for each light curve. After this procedure, a timing error larger than $\sim 0.33\,\mathrm{s}$ is unlikely to persist.

For the further analysis, two versions of the total light curve were used. First, the original, non-weighted light curve was used for prewhitening (see Sect.~\ref{non-weighted_section}). Then, weights were estimated calculating a moving standard deviation of the residual light curve. In the overlapping regions, only the light curve data set with the highest weights was retained.

\section{Prewhitening}	\label{prewhitening_section}
\subsection{Non-weighted light curve}	\label{non-weighted_section}
Using the program \verb|Period04| by P. Lenz and M. Breger, the original, non-weighted light curve was prewhitened until residual peaks fell below a limit of four times the noise level. First, the total light curve was divided into two roughly equal parts with a time basis of 16.0\,d and 14.6\,d respectively, leading to a frequency resolution $\Delta f \sim 0.062\,$c$\,$d$^{-1}$ and $\sim 0.068\,$c$\,$d$^{-1}$. These light curves were prewhitened individually in order to look for frequencies present in both halves of the data set. Due to the inhomogeneity of the photometry, 75 frequencies could be fitted for the first half's data, but only 41 for the second half \citep[see Langfellner et~al., in preparation, or][for details]{langfellner_2011}. As a result, only 30 frequencies were found to be present in both halves.

Prewhitening of the total light curve with a frequency resolution of $\sim 0.033\,$c$\,$d$^{-1}$ yielded 85 frequencies in total, situated in the range between 150 and 550\,c$\,$d$^{-1}$. For higher frequencies, no significant frequencies were spotted. For frequencies below $\sim$100\,c$\,$d$^{-1}$, pulsation frequencies become undistinguishable from atmospheric effects. 

\begin{figure}[!ht]
 \centering
\includegraphics[angle=90,width=\textwidth,clip=true]{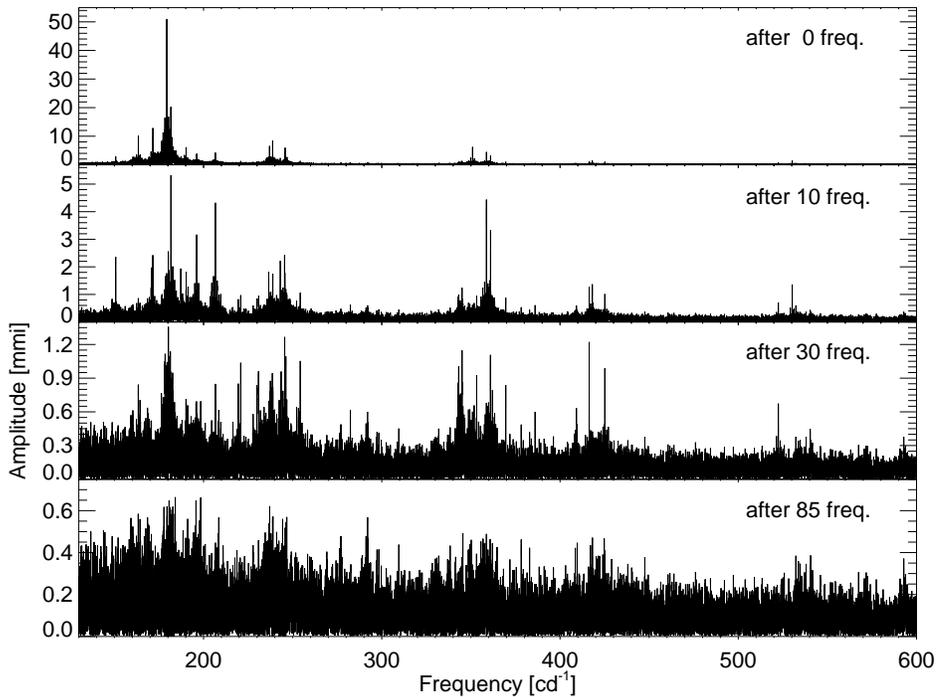}
\caption{Prewhitening process of the original, non-weighted light curve. The uppermost panel shows the original Fourier spectrum, the panels below the residual spectra after prewhitening 10, 30 and 85 frequencies.}	\label{prewhitening_fig}
\end{figure}

Snapshots of the prewhitening process are shown in fig.~\ref{prewhitening_fig}. The 10 frequencies with the highest amplitudes are listed in table~\ref{frequency_table}. Very close to the main mode, another strong peak was detected. As the frequency difference is far smaller than the resolution and the second peak had not been detected in the split light curves, we assume this peak to be artificial and give it the rank ``1a''. The combination of $f_1$ and $f_\mathrm{1a}$ should result in an amplitude near 50\,mmi, as visible in fig.~\ref{prewhitening_fig}.

Out of the 85 frequencies, 18 frequency sums and the first harmonics of the strongest three modes could be found with deviations $\delta f$ well below the frequency resolution. Remarkably, all combinations of the main mode with the other strong modes up to rank 16 are present in the data, apart from ranks 13 and 15. Furthermore, the three highest frequencies consist of the sums of the first harmonics of $f_1$ or $f_3$ and another strong frequency. This is indicative of non-linear pulsations \citep{antonello_1998}.

\begin{table}[!ht]
\caption{The ten strongest frequencies in MSST/WET non-weighted light curve. Note the splitting of the main mode $f_1$ into two frequencies with ranks ``1'' and ``1a''. The rank of $f_{10}$ is set in brackets, as this frequency could not be resolved in both halves of the light curve. See Langfellner et~al.\ (in preparation) for a complete frequency table.}	\label{frequency_table}
\smallskip
\begin{center}
{\small
\begin{tabular}{cccccc}
\tableline
\noalign{\smallskip}
Rank	&	$f$ 	&	$A$	&	$P$ 	&	Freq. sum	&	$\delta f$	\\
	&	[$\mathrm{c}\,\mathrm{d}^{-1}$]	& [mi]	& [s]	& &	[$\mathrm{c}\,\mathrm{d}^{-1}$] \\
\noalign{\smallskip}
\tableline
\noalign{\smallskip}
4	&	163.415	&	0.0101	&	528.71	&		&		\\
3	&	171.574	&	0.0121	&	503.57	&		&		\\
1a	&	179.343	&	0.0205	&	481.76	&		&		\\
1	&	179.347	&	0.0668	&	481.75	&		&		\\
6	&	181.608	&	0.0067	&	475.75	&		&		\\
2	&	181.659	&	0.0151	&	475.62	&		&		\\
(10)	&	181.703	&	0.0055	&	475.50	&		&		\\
8	&	236.997	&	0.0062	&	364.56	&		&		\\
5	&	238.752	&	0.0081	&	361.88	&		&		\\
9	&	245.781	&	0.0058	&	351.53	&		&		\\
7	&	350.920	&	0.0062	&	246.21	&	$f_1+f_3$	&	0.001	\\
\noalign{\smallskip}
\tableline
\end{tabular}
}
\end{center}
\end{table}

\subsection{Comparison with Kilkenny et~al.\ (1999)}
Let us compare the results with those from \citet{kilkenny_1999}, representing the most extensive previous photometric multi-site campaign on PG\,1605+072. Fortunately, at least 32 of the total 55 frequencies match the MSST/WET results, i.\,e.\ they differ by less than $\Delta f$. At least one additional frequency match could be explained by aliasing ($180.496\,\mathrm{c}\,\mathrm{d}^{-1}$ in MSST/WET and $181.475\,\mathrm{c}\,\mathrm{d}^{-1}$ in \citeauthor{kilkenny_1999}\ data). However, there are also stronger frequencies in the MSST/WET data, such as $f_5$, that did not have any corresponding frequency detection.

As expected, the MSST and WET campaigns, covering twice the time of the \citeauthor{kilkenny_1999}\ campaign, deliver more significant frequencies -- 85 compared to 55. However, the number of frequencies confirmed by appearing in both halves of the data is smaller (22 vs.\ 28) due to the inhomogeneity of the MSST/WET data. Besides, the number of frequencies that can be constructed by the sum of two other frequencies has almost doubled -- from 10 to 18. Additionally, the first harmonics can be found for the three strongest frequencies instead of just for the main mode.

It is remarkable as well that the frequency range could be extended by identifying three significant frequencies in the region between 500 and $600\,\mathrm{c}\,\mathrm{d}^{-1}$. Thus, the pulsation periods reach down at least to 160\,s. However, higher harmonics had already been found by \citet{koen_1998}, reaching up to the third harmonic of $f_2$.

Apart from the frequency values, the comparison of the amplitudes gives interesting results. The main mode is much stronger in the MSST/WET data, even if the two frequencies ranked ``1'' and ``1a'' are thought to interfere destructively. This frequency splitting in the Fourier spectrum is another indication for amplitude variation. 

Another remarkable result is the frequency triplet comprising $f_6$, $f_2$ and $f_{10}$ (see table~\ref{frequency_table}), as it is also identifiable in the \citeauthor{kilkenny_1999}\ data, but with completely different amplitudes. In general, the amplitude values change a lot -- a frequency ranked as high as ``5'' in the MSST/WET data even does not appear at all in the old data.

\subsection{Using weights}	\label{weight_section}
As mentioned in Sect.~\ref{data_section}, the residuals of the non-weighted prewhitening process were used to derive point weights that served as a criterion for choosing the best light curve fragment in overlapping regions. For the weight derivation, a \verb|Fortran| program by Roberto Silvotti was used that calculates a moving standard deviation of the residuals.

This modified, weighted light curve was used to repeat the prewhitening process. As before, 85 frequencies could be identified. Taking the frequency resolution into account, we found that 71 of the frequencies are identical to the ones from the non-weighted prewhitening procedure. Four more frequency sums could be identified, yielding a total of 22. Again, we found the first harmonics of the three strongest modes.

Considering the main mode, its double feature disappeared and showed only a single peak with an amplitude reduced to $\sim 51\,$mmi. The amplitudes of the other 15 strongest frequencies only changed slightly (i.\,e.\ $\lesssim 10\%$). Merely for amplitude ranks $\gtrsim 20$, particularly $\gtrsim 60$, frequencies that were not found in the non-weighted light curve were spotted. In general, however, these differences are not substantial. Consequently, the non-weighted light curve could be used for the further analysis.

\section{Amplitude variation}	\label{amp-variation_section}
\begin{figure}[!ht]
 \centering
\includegraphics[width=\textwidth,clip=true]{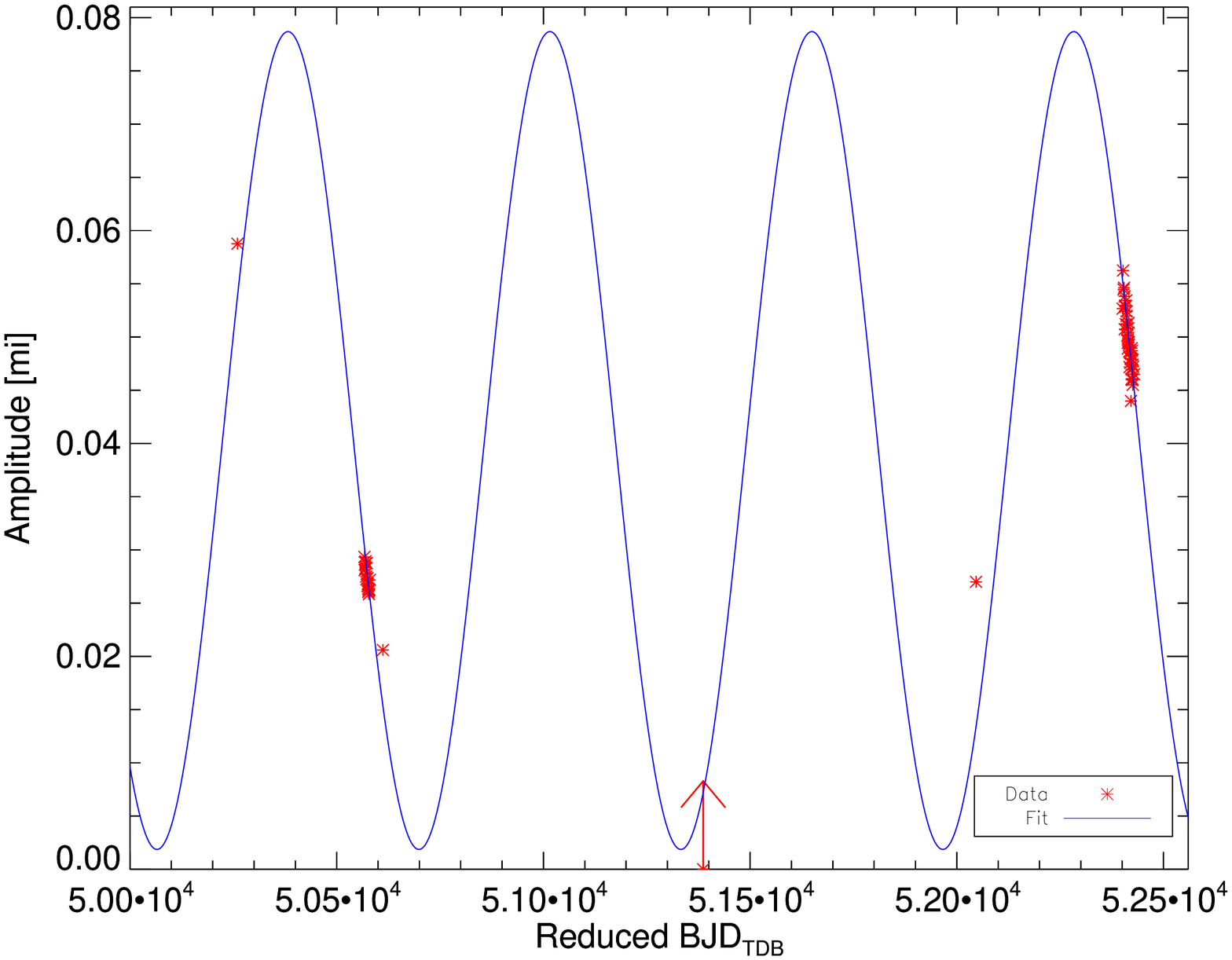}
\caption{Long-term beating of main mode $f_1$. Together with the 40 amplitude values from O$-$C diagrams for $f_1$ in the MSST/WET data (right-hand side of the plot) and the 20 amplitude values from analysing \citet{kilkenny_1999} data (left-hand side), mean amplitude values from previous campaigns are plotted, appearing as single stars. These are taken from (in chronological order) \citet{koen_1998}, \citet{spaandonk_2008}, \citet{otoole_2000} and \citet{falter_2003}. Note, however, that for the sinusoid fit that is also shown in the plot the original mean amplitude $\sim 0.0274$\,mi for the campaign by \citet{kilkenny_1999} was used. In the data from \citet{otoole_2000}, $f_1$ could not be found at all. Therefore, the amplitude value was set to zero. However, $f_1$ could have had an amplitude bigger than zero that was hidden in the noise. This is indicated by the arrow. Since the amplitude value sensitively depends on possible unresolved beating or destructive interference with noise peaks, values for the amplitude error are not given.}	\label{beating_longterm_fig}
\end{figure}

For the search for amplitude variation visible on the timescale of the MSST/WET observations, the O$-$C diagram with its high sensitivity proved to be the tool of choice. For the main mode, we divided the non-weighted total light curve into 40 chunks, each consisting of a comparable amount of data points, and fitted the main mode's amplitude and phase for each part. The ``ghost'' frequency $f_\mathrm{1a}$ was excluded from the fit, as we consider it to be an artefact from imperfectly overlapping data sets (see Sect.~\ref{weight_section}) or amplitude variation. The result was a linear decreasing trend in the amplitude as well as the phase. The latter trend could be fixed by adding $0.00138\,\mathrm{c}\,\mathrm{d}^{-1}$ to the initial frequency guess $f_\mathrm{guess} \approx 179.34699\,\mathrm{c}\,\mathrm{d}^{-1}$, delivering the corrected value $f_1 \approx 179.34837\,\mathrm{c}\,\mathrm{d}^{-1}$, while the linear amplitude decrease proved to persist at a rate of $(-0.3395\pm0.0089)\,$mmi\,d$^{-1}$.

In order to put this result into a wider context, we compared the main mode's amplitude value and change rate with previous photometric investigations. A plot of the amplitudes versus time is shown in fig.~\ref{beating_longterm_fig}. See the caption for the origin of the data points. As the simplest reason for amplitude variation is beating between two close unresolved frequencies, we tried to fit the data with a sinusoid function. The result with a period of $\sim 630\,$d matches the data remarkably well.

As a further test, we looked for amplitude variation in the data from \citet{kilkenny_1999}, applying the same method as for $f_1$ in the MSST/WET data, but using only 20 chunks. As $f_{48}$ in those data is closer to $f_1$ than the frequency resolution ($\sim 0.07\,\mathrm{c}\,\mathrm{d}^{-1}$), it was excluded prior to the analysis. Apart from sinusoidal beating between $f_1$ and $f_{48}$, the result was a linear decrease in the amplitude, similar to the MSST/WET data, but with a smaller slope. Since the proposed sine indeed expresses a shallower slope there, this supports the beating hypothesis.

Notably, a linear decrease with a slope twice as steep could be found in the MSST/WET data for $f_{13}$, which is the first harmonic of $f_1$.

\section{Multiplet structure}	\label{multiplet_section}
The frequency triplet consisting of $f_6$, $f_2$ and $f_{10}$ (see table~\ref{frequency_table}) does also exist in the data from \citet{kilkenny_1999}, however with a different amplitude distribution: Whereas the side peaks both had an amplitude of $\sim 15\,$mmi before (putting them on ranks ``2'' and ``3''), they weakened to $\sim 6\,$mmi in the MSST/WET data. The central peak -- now $f_2$ -- was barely detectable as $f_{32}$ with an amplitude of $\sim 1\,$mmi. Furthermore, the splitting of the triplet reduced from $\sim 0.07\,\mathrm{c}\,\mathrm{d}^{-1}$ to $\sim 0.04\,\mathrm{c}\,\mathrm{d}^{-1}$, both being in the area of the respective frequency resolution. This implies that the true splitting of the triplet is even closer and probably has remained constant rather than having changed over five years (Suzanna Randall, private communication).

\begin{figure}[!ht]
 \centering
\includegraphics[angle=90,width=\textwidth,clip=true]{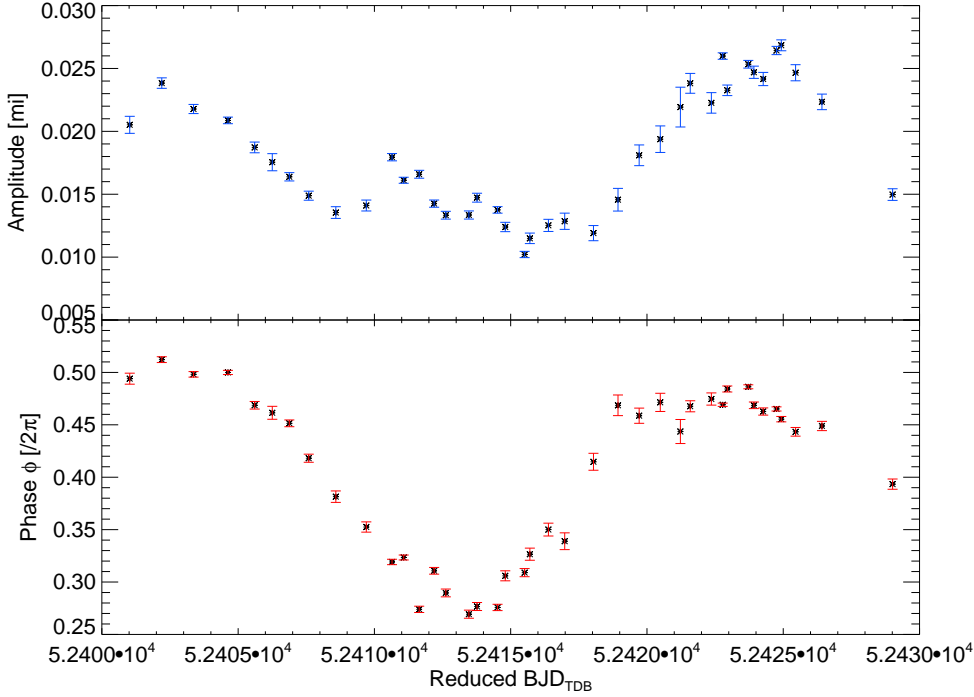}
\caption{O$-$C diagram of $f_2$, $f_6$ and $f_{10}$ when only considering $f_2$. The amplitude and phase vary like a sine with the same period, which is indicative of beating.}	\label{beating_F2_fig}
\end{figure}

On the other hand, it is also possible that the triplet structure is an artefact appearing due to amplitude variation. However, \citet{guggenberger_2008} demonstrated that the triplet (or effectively a doublet due to the weak central frequency) in the data from \citet{kilkenny_1999} indeed is real.

To test if the ``new'' triplet is real, we excluded $f_6$ and $f_{10}$ from the frequency set and checked $f_2$ for amplitude variation. The result (see fig.~\ref{beating_F2_fig}) shows a strong indication for a sine-like periodic function, with both the same period in the amplitude and phase. Although beating involving three frequencies is far more complicated than beating between two frequencies, the result is clear enough to support the hypothesis.

In addition to this triplet, other multiplet candidates with a similar spacing could be found in the (complete) frequency table, also showing signs indicative of beating (see Langfellner et~al., in preparation).

\section{Discussion}	\label{discussion_section}
\subsection{PG\,1605+072: a slow rotator?}
The analysis of the triplet comprising $f_2$, $f_6$ and $f_{10}$ indicates that these frequencies are real, although the exact frequency spacing is not known due to the limited frequency resolution. However, at least an upper limit can be given with $\sim 0.05\,\mathrm{c}\,\mathrm{d}^{-1}$.

The most plausible explanation for the existence of the multiplets is rotational splitting, lifting the $m$ degeneracy. For solid, slow ($P_\mathrm{rot} \gtrsim 2.5\,\mathrm{h}$) rotators and considering $p$-modes, the frequency spacing is governed by $P_\mathrm{rot} \sim 1/\Delta f_{kl}$ \citep{charpinet_2008}, including the rotation rate $V_\mathrm{rot} = 1/P_\mathrm{rot}$ and the frequency spacing $\Delta f_{kl}$, depending on the radial order $k$. Assuming the upper limit for the frequency spacing of $\sim 0.05\,\mathrm{c}\,\mathrm{d}^{-1}$, a minimum rotation period of about $20\,$d is inferred. Using crude estimates and considering both the high-mass and evolved status scenarios, the equatorial rotation velocity is then found to be $V_\mathrm{eq} \lesssim 0.9\,$km$\,$s$^{-1}$ or $\lesssim 1.3\,$km$\,$s$^{-1}$ respectively.

This disagrees with the claims of \citet{heber_1999} and \citet{kawaler_1999} that have been discussed in Sect.~\ref{introduction_section} by orders of magnitude. However, they assumed rotational broadening or non-trapped modes. Since \citet{kuassivi_2005} introduced the possibility of pulsational broadening, which was confirmed by \citet{otoole_2005}, it is very possible that PG\,1605+072 is a slow rotator.

Theoretical models by \citet{spaandonk_2008} and \citet{vangrootel_2010} cover multiplet spacings of $\sim 3.5\,\mathrm{c}\,\mathrm{d}^{-1}$ or $\sim 7.8\,\mathrm{c}\,\mathrm{d}^{-1}$ ($\sim 40\,\mu$Hz or $\sim 90\,\mu$Hz, meaning fast rotation) as well as slow-rotation models. Indeed, the reasonable assumption that the main mode is radial -- as it is part of no multiplet with a spacing around $\sim 0.05\,\mathrm{c}\,\mathrm{d}^{-1}$ -- constrains the available models sufficiently well to leave a compatible slow-rotation model by \citet{spaandonk_2008}, implying a high mass $M \sim 0.7\,M_\odot$. Since \citet{vangrootel_2010} came to the conclusion that all the models they presented predicted a mass $M \sim 0.76\,M_\odot$, this not only supports that PG\,1605+072 is a slow rotator, but also that it possesses a high mass rather than an evolved status.

\subsection{Understanding the main mode's amplitude variation}
The sine fitted to the main mode's amplitude data from various campaigns is able to describe the position of the data points as well as the slope in the data from the MSST/WET and \citet{kilkenny_1999} campaigns, making us believe it to be correct.

However, the question then arises of why the main mode varies with such a long period. For beating, a very close frequency doublet would be required. If the multiplets discussed in the previous section are due to rotational splitting, the long-term beating must have another reason. If, on the other hand, the long-term beating would be treated as being due to rotational splitting itself, extremely low rotation rates in the order of $0.03\,\mathrm{km}\,\mathrm{s}^{-1}$ would be the result, which seems far-fetched. Should the frequency doublet be caused by another physical effect, it is -- so far -- unclear what specific mechanism is applicable.

\section{Summary and conclusions}	\label{conclusions_section}
The prewhitening of the combined MSST and WET XCov22 photometric data delivered 85 identifiable frequencies, with merely small deviations if a weighted light curve was used. At least 21 of these frequencies are either harmonics or frequency sums. A comparison with the results from \citet{kilkenny_1999}, representing the most extensive photometric data on PG\,1605+072 prior to MSST/WET, yielded 32 common frequencies out of 55. In total, the MSST and WET data increase the number of simultaneously identified frequencies by 30.

The main mode's amplitude was found to decrease linearly on the timescale of the observations. Putting this result in the context of the previous campaigns, we found that the amplitude variation can be described by a sinusoid implying long-term beating with a period of $\sim 630\,$d. The validity of the sine fit was supported by comparing its slope to the amplitude variation rate in the data from \citet{kilkenny_1999}.

We found that the frequencies with the amplitude ranks ``2'', ``6'' and ``10'' form a real triplet, which has also been present in the data from \citet{kilkenny_1999}, yet changed its amplitude values. Furthermore, the observed frequency spacings have become smaller, reflecting the better frequency resolution. This indicates that the real spacings might be still closer. Additionally, similar multiplet features were found for many other frequencies. A more detailed description will be available in Langfellner et~al.\ (in preparation).

The multiplet structure is most likely caused by rotational splitting, indicative of PG\,1605+072 being a slow rotator. This is in agreement with studies of white dwarfs and other single sdB stars. A comparison with theoretical asteroseismic models supports the slow-rotation hypothesis and favours the high-mass scenario instead of an evolved status. Consequently, this would require an alternative formation channel.

These results help to further constrain the possible features of PG\,1605+072. The photometric data and the extensive frequency tables will be published in Langfellner et~al.\ (in preparation) and made available online for further investigation.

\acknowledgements We acknowledge and thank the MSST and WET XCov22 observers for their efforts in obtaining the photometric data used in this analysis. We would further like to thank Dave Kilkenny for kindly providing the 1997 campaign data. Jan Langfellner would like to thank the sdOB5 conference LOC for kindly waiving the conference fee.

\bibliography{literature.bib}

\end{document}